\documentclass[prl,twocolumn,nofootinbib]{revtex4}
\pdfoutput=1 
\usepackage{amsmath,amssymb}
\usepackage{hyperref}
\usepackage{graphicx}
\usepackage{multirow}
\usepackage{color}
\usepackage[utf8]{inputenc}

\def\beq{\begin{equation}}
\def\eeq{\end{equation}}
\def\beqa{\begin{eqnarray}}
\def\eeqa{\end{eqnarray}}
\def\bit{\begin{itemize}}
\def\eit{\end{itemize}}

\begin{document}
\title{\boldmath{$J_{E_T}^{\rm II}$:} A Two-prong Jet Finding Algorithm}
\author{
Yang Bai,$^{a}$ Zhenyu Han$^{b}$ and Ran Lu$^{a}$
\vspace{5mm}
\\
$^{a}$ \normalsize\emph{Department of Physics, University of Wisconsin-Madison,  Madison, WI 53706, USA}  \vspace{1mm} \\
$^{b}$ \normalsize\emph{Institute for Theoretical Science, University of Oregon, Eugene, OR 97403, USA}
}
\begin{abstract}
We propose a new global jet-finding algorithm for reconstructing two-prong objects like hadronic weak gauge bosons at a hadron collider. The selection of particles in a two-prong jet is required to maximize a $J_{E_T}^{\rm II}$ function, which contains a modified second Fox-Wolfram moment and prefers a two-prong structure for a fixed jet mass. Compared to the traditional jet-substructure method, our algorithm can provide a similar or better performance for identifying boosted weak gauge bosons that are produced from either Standard Model processes or heavy di-boson resonance decays.
\end{abstract}
\maketitle
\noindent
{\it{\textbf{Introduction.}}}
Jet substructure techniques have become essential tools in the search for new physics at a hadron collider such as the Larger Hadron Collider (LHC). Strategies have been proposed to reconstruct hadronically decaying objects such as the $W$ and $Z$ gauge bosons, the Higgs boson and the top quark, and extract from them vital information that was perviously unattainable. Most existing jet substructure methods are separated into two steps: first, jets are formed with standard jet-finding algorithms~\cite{Ellis:1993tq,Salam:2009jx}. The jet size must be large enough to include most of the decay products of the object. Therefore the jets are ``fat''. Fat jets are then processed with a grooming algorithm~\cite{Butterworth:2008iy, Ellis:2009su,Ellis:2009me, Krohn:2009th,Larkoski:2014wba} or other methods to remove soft particles that are unlikely from the object decay. An alternative, relatively unexplored approach is to develop a new jet-clustering algorithm to directly reconstruct the multi-prong objects, utilizing its structure and properties.~\footnote{The $N$-jettiness/XCone~\cite{Stewart:2010tn, Stewart:2015waa, Thaler:2015xaa} method can be viewed as such an attempt, although in which a fixed number of jets in an event are constructed.} The algorithm could be more specialized to suit different objects, and, if carefully designed, should be more efficient to reconstruct the physical object than ordinary jet-finding algorithms because it contains more physics information.

To reconstruct a single-prong jet, one can choose a set of particles to maximize a jet function that prefers a large jet momentum but a small jet mass~\cite{Georgi:2014zwa,Ge:2014ova,Bai:2014qca,Kaufmann:2014nda}.~\footnote{See Ref.~\cite{Thaler:2015uja} for its connection to other jet finding algorithms.} The relevant jet function for one-prong jet only depends on the summed four-momenta of particles belonging to the jet. Following the same philosophy, in this paper we develop a new jet-finding algorithm to reconstruct two-prong jets. Since the two-prong structure can be interpreted as a jet shape, the immediate question is how to add an additional component in the jet function to prefer a two-prong structure. There is a large amount of literature on event shapes at a lepton collider including the Fox-Wolfram (FW) moments~\cite{Fox:1978vu,Fox:1978vw}. We will show that with some modifications, we can incorporate the second Fox-Wolfram moment into the jet function, maximizing which prefers a two-prong structure for a jet. 

Our new jet-finding algorithm is a global one because all subsets of particles will be considered to maximize the jet function. It has advantages of including only relevant particles for the physical object, and being infrared and collinear safe based on the argument in Ref.~\cite{Georgi:2014zwa} and the discussion on FW moments in Refs.~\cite{Fox:1978vu,Fox:1978vw}. On the other hand, it also has its disadvantage because $2^N$ subsets need to be checked, where $N$ is the number of particles in an event, and an unrealistic computing power is required for just dozen of particles. To reconcile this problem, we provide a double-anti-$k_t$-cone algorithm to find the approximate maximum of the jet function. As we will demonstrate, this approximation matches very well the true global maximum. As a result, the double-anti-$k_t$-cone algorithm has a realistic number of operations of $\mathcal{O}(N^2 n_R^3)$, where $n_R$ is the number of different anti-$k_t$ jet sizes. 

\noindent
{\it{\textbf{Two-prong Jet Function.}}}
Similarly to the studies in Refs.~\cite{Georgi:2014zwa,Ge:2014ova,Bai:2014qca} for a one-prong jet, we maximize the following two-prong jet function to select the subset of particles in an event as a jet
\beqa
J_{E}^{\rm II} &=& E^2 - \beta\,M^2 +\gamma\,\widetilde{H}_2  \quad \mbox{(lepton colliders)}  
\label{eq:jet-lepton} \,, \\
J_{E_T}^{\rm II} &=& E_T^2 - \beta\,M^2 + \gamma\,\widetilde{H}_2 \quad \mbox{(hadron colliders)} 
\label{eq:jet-hadron} \,. 
\eeqa
Here, $E$ and $M$ are the jet energy in the lab frame and jet mass for the subset of (massless) particles. For hadron colliders, we use the transverse energy defined as $E_T^2 \equiv E^2 - P_z^2$. In addition to the one-prong jet function~\cite{Bai:2014qca}, we have added an additional component,
\beqa
    \widetilde{H}_2= M^2\Big[\sum_{i,j}\frac{\left( p_i \cdot p_j \right)^2}{\left( P_J \cdot p_i \right)\left( P_J \cdot p_j \right)} - 1\Big] \,,  \label{def:h2t_l}
\eeqa
with $P_J^\mu \equiv \sum_i p_i^\mu$ the total four-momentum of the subset and $p_i \cdot p_j = p_i^\mu p_{j \mu}$. 

A simpler way to understand this new function is to boost all particles to the rest frame of $P_J^\mu$, in which one has
\beqa
\widetilde{H}_2 \equiv \Big(\sum_{i,j} E_i E_j \cos^2\theta_{ij} \Big)_{\mbox{\footnotesize rest frame}}   \,.
\label{def:h2t}
\eeqa
Here, $\theta_{ij}$ is the angle between two particles in the rest frame of $P_J^\mu$. For a fixed invariant mass $\sum_i E_i^{\mbox{\footnotesize rest\,frame }} = M$, one can easily show that $\widetilde{H}_2$ is maximized when there are only two back-to-back massless prongs. Our function is related to the FW moments~\cite{Fox:1978vu,Fox:1978vw}, $H_{l} \equiv \sum_{i, j} \left\lvert\vec{p}_i \right\rvert \left\lvert\vec{p}_j\right\rvert P_l(\cos\theta_{ij})/E^2$, by $\widetilde{H}_2 = E^2 ( 2 H_2 + 1)/3$. It is also interesting to note that the zeroth and first FW moments are equivalent to the jet energy and mass, respectively. Therefore, our $J_E^{\rm II}$ can be expressed just using $H_{0,1,2}$.

For the one-prong jet function in Refs.~\cite{Georgi:2014zwa,Ge:2014ova,Bai:2014qca}, it was shown that the jet parameter $\beta$ controls the size of the jet with the jet cone radius scaling as $1/\sqrt{\beta}$ for a large $\beta$. Our two-prong jet functions in Eqs.~(\ref{eq:jet-lepton}) and (\ref{eq:jet-hadron}) have two parameters, $\beta$ and $\gamma$. Similar to the one-prong case, $\beta$ should be positive to avoid clustering all particles in an event into a single jet. Furthermore, we also require $\beta \geq \gamma$, otherwise two back-to-back groups of particles are preferred in a single jet. This can be easily seen for two massless energetic particles, which have $J_{E}^{\rm II} = E^2 - \left( \beta - \gamma \right) M^2$ and $(\beta-\gamma)$ as the effective $\beta$ for the one-prong jet function. Therefore, we anticipate that the parameter $1/\sqrt{\beta-\gamma}$ for a large $(\beta-\gamma)$ roughly controls the geometric size of a fat two-prong jet. At hadron colliders, in order to exclude particles in the beam direction, we also require $\beta -\gamma > 1$.

\begin{figure}[thb!]
    \centering
    \includegraphics[width=0.45\textwidth]{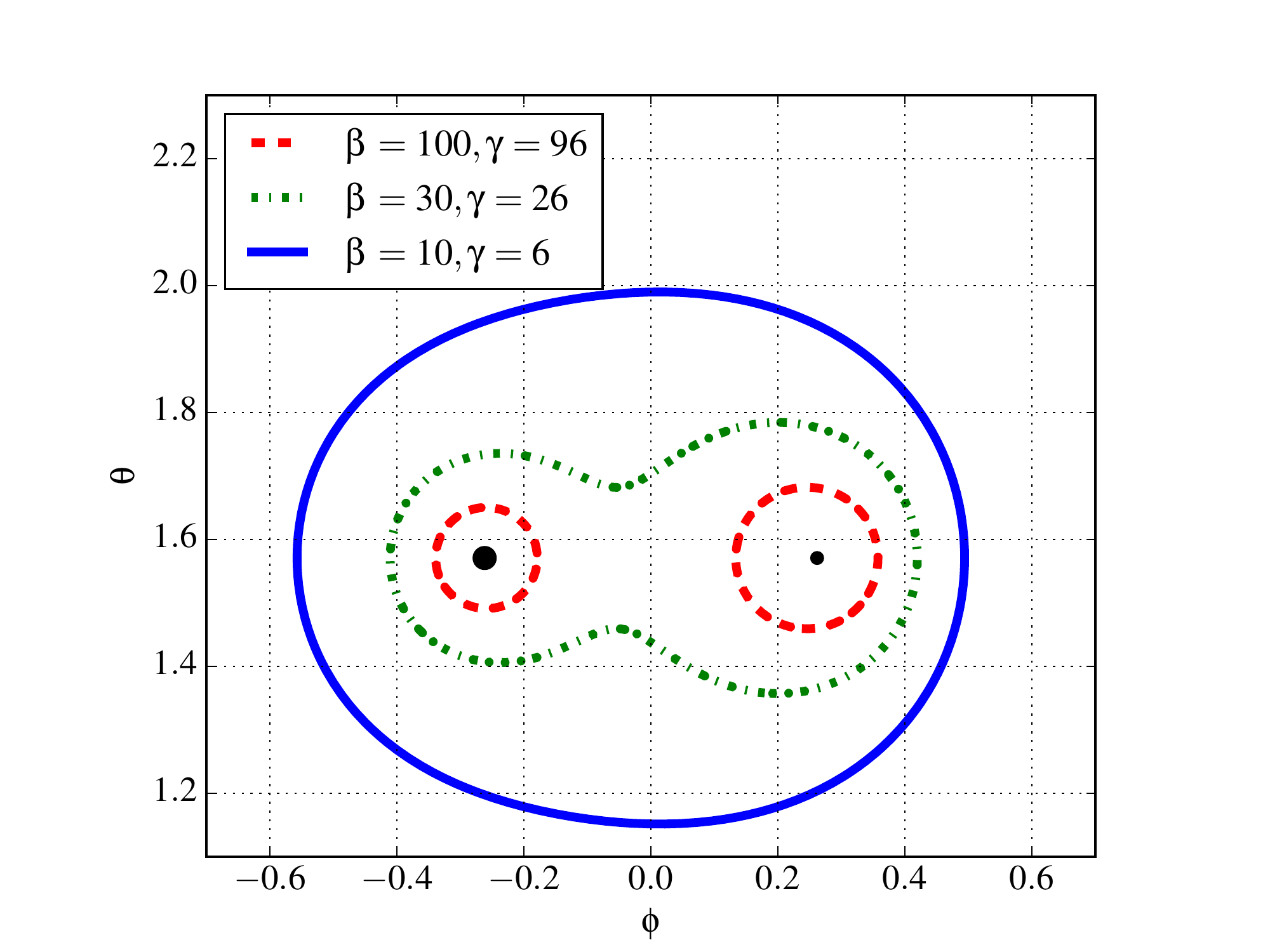}
    \caption{The boundaries of soft particles surrounding two hard particles with $E_1 = 3 E_2$ on the $\theta-\phi$ plane for different values of $\beta$ and $\gamma$. The locations of the two hard particles are marked with the black dots.}
  \label{fig:boundary}
\end{figure}

To further understand the shape of an object obtained by maximizing the two-prong jet function, we add an infinitely soft particle into the system of two energetic massless prongs to determine the passive catchment area~\cite{Cacciari:2008gn}. Since the generic feature at a hadron collider is similar to the one at a lepton collider, we study the jet function at a lepton collider for simplicity. Defining the weighted angular distances of the soft particle with respect to the hard particles
\beqa
 d_1 = \frac{2\,E_1\,(1-\cos\theta_{1})}{E_1+E_2}\,, \quad
 d_2 = \frac{2\,E_2\,(1-\cos\theta_{2})}{E_1+E_2}\,,   \label{def:dist}
\eeqa
with $\theta_1$ and $\theta_2$ as the angles between the soft particle and the two hard particles, we derive the following condition for the soft particle to increase the value of $J_E^{\rm II}$
\beqa
\beta ( d_1 + d_2)^2 - \gamma (d_1 - d_2)^2 < 2 (d_1 + d_2) \,.
    \label{ineq:cond}
\eeqa
Choosing the two hard particles with $E_1 = 3 E_2$ located at $(\pi/2, \pm \pi /12)$, we show the boundary of soft particles in Fig.~\ref{fig:boundary} for different values of $\beta$ and $\gamma$. 
From Eq.~(\ref{ineq:cond}) and Fig.~\ref{fig:boundary}, one can see that for a large value of $\beta$ the soft particles are close to one of the two hard particles and the allowed region consists of two separated cones. As $\beta$ decreases, the two cones expand and overlap with each other. For a small value of $\beta$, the ``fat jet" has a single cone with two subjets inside, which is what the traditional jet-substructure method searches for.

\noindent
{\it{\textbf{Double-anti-$k_T$-Cone Algorithm.}}}
It is unrealistic to exhaust all $2^N$ possible subsets of particles to maximize our jet function, even for several dozen of particles in an event. In practice, however, one can find approximate and fast ways to identify a subset of particles that provide a jet function value close to the true maximum. We provide one such algorithm built on top of the anti-$k_t$ algorithm~\cite{Cacciari:2008gp}.  

Given the observation in Fig.~\ref{fig:boundary} that the jet shape can be described roughly by two separate or overlapping cones, one could consider all possible pairs of cones (and individual ones), and identify the one with the largest $J_{E_T}^{\rm II}$ as the jet candidate. Specifically, our jet-finding algorithm is described as following
\begin{enumerate}
\item Preparing candidate cones: for a given event, collect anti-$k_t$ (or other fast cone-shape-jet-finding algorithm) jets with various jet radii. For instance, one can scan the jet radii in $n_R\sim{\cal O}(10)$ steps with a maximum value of $R_{\rm max}\sim{\cal O}(1)$. In the following simulation, we will choose $R_{\rm max}=0.7$ and $R_{\rm min}=R_{\rm max}/n_R = 0.05$. 
\item Maximizing $J_{E_T}^{\rm II}$: for each pair of (and individual) anti-$k_t$ jets, calculate $J_{E_T}^{\rm II}$ based on the particles belong to this pair of (or individual) anti-$k_t$ jets. Particles shared by two different jets are counted only once. Find the pair (or the individual one) that maximizes $J_{E_T}^{\rm II}$ as the leading jet. 
\item Iterating: remove particles that belong to the leading jet, and repeat the procedure to find the next jet, until all particles in an event are exhausted. 
\end{enumerate}
The running time of this algorithm for $N$ particles scales as $N^2 n_R^3$.~\footnote{Our numerical code can be found in \url{https://github.com/LHCJet/JETII}, which is based on \texttt{FastJet}~\cite{Cacciari:2011ma}.} We test the goodness of the above approximate algorithm against the true maximum using a sample of hadronic $WW$ events generated with a $p^{W}_{T}>200$ GeV cut. Since the true maximum requires us to exhaust all $2^N$ possible subsets of particles, we only keep the leading 20 particles in our test. We have found that over 99\% of the events, the double-anti-$k_T$-cone algorithm finds exactly the same set of particles for the leading jet as the exhaustive subset method. For the remaining less than one percent of events, the particle contents only differ slightly with the difference of the jet $p_T$'s less than 10\%.

\noindent
{\it{\textbf{Results for W-jet Tagging.}}}
Using boosted $W$-bosons as an example, we compare the results from our $J_{E_T}^{\rm II}$ jet-finding algorithm with other traditional jet-substructure methods including the jet grooming methods~\cite{Butterworth:2008iy, Ellis:2009su,Ellis:2009me, Krohn:2009th}. For the signal, we consider the Standard Model $WW$ pair production at the 14 TeV LHC with both $W$'s decay hadronically. The dominant background is QCD dijets. Both signal and background events are simulated with \texttt{PYTHIA\,8}~\cite{Sjostrand:2014zea}. To mimic the pileup effects, we also add 20 soft QCD events for each hard event.  

\begin{figure}[thb!]
    \centering
    \includegraphics[width=0.235\textwidth, clip=true,viewport= 15 0 540 400]{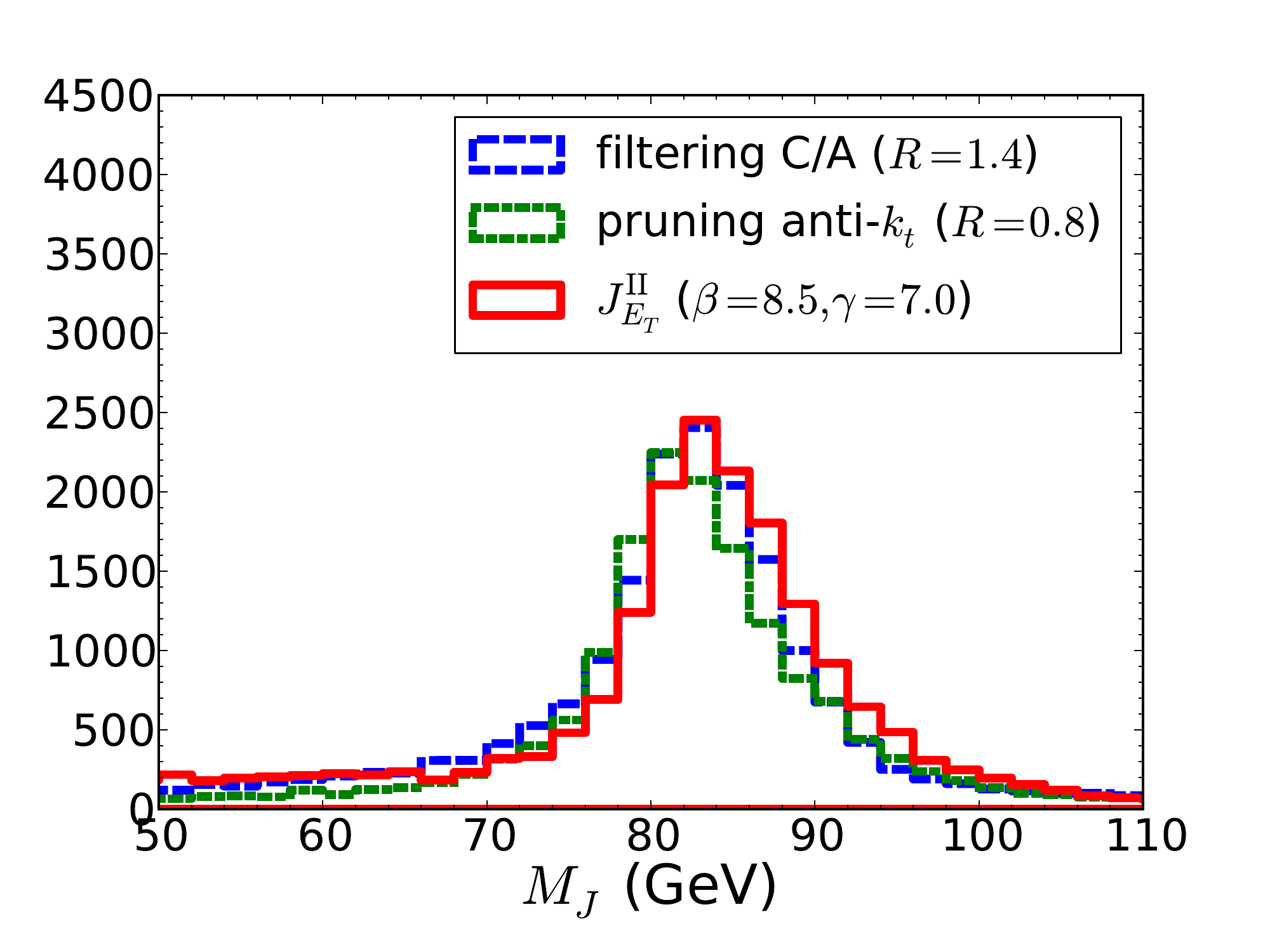} \hspace{0mm}
    \includegraphics[width=0.235\textwidth, clip=true,viewport= 5 0 530 400]{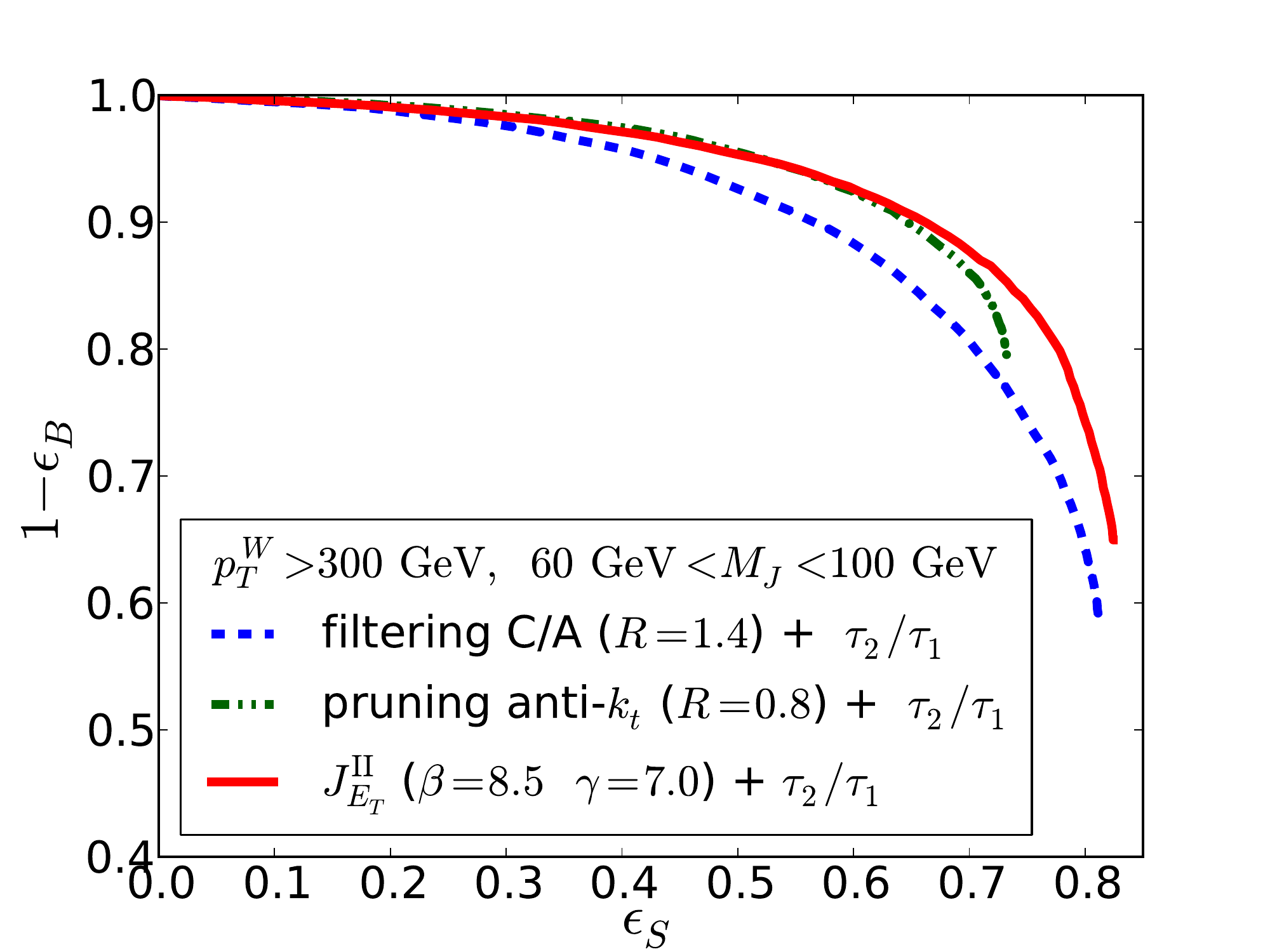}
    \caption{Left panel: jet mass distributions for the leading two jets. The jet algorithm parameters for obtaining the largest signal efficiency can be found in Table~\ref{tab:params} for $p_T^W > 300$~GeV. Right panel: the receiver operating characteristic curves for the signal and background efficiencies by varying the cut on the $N$-subjettiness ratio variable $\tau_{2}/\tau_{1}$~\cite{Thaler:2010tr}.}
   \label{fig:hist}
\end{figure}

Imposing a parton-level $p_T > 300$~GeV cut on the $W$'s and QCD jets, we show the reconstructed jet mass distributions in the left panel of Fig.~\ref{fig:hist}, comparing our jet-finding algorithm and the traditional jet-substructure analysis with a fat jet plus filtering or pruning. We have chosen to maximize the signal acceptance for jet mass in the $(60, 100)$~GeV window, and the optimum jet parameters are shown in Table~\ref{tab:params}. From the left-panel of Fig.~\ref{fig:hist}, one can see that our jet-finding algorithm identifies more $W$-jets in the mass window. For events within the $(60, 100)$~GeV mass window, we further calculate the $N$-subjettiness ratio, $\tau_2/\tau_1$~\cite{Thaler:2010tr}. We apply a varying cut on $\tau_{2}/\tau_{1}$ to produce the receiver operating characteristic (ROC) curves, shown in the right panel of Fig.~\ref{fig:hist}.  Comparing to filtering and pruning, our jet algorithm provides a comparable or larger signal acceptance for most of the background efficiencies. The ROC curve of pruning is very close to the $J_{E_T}^{\rm II}$ one when the signal efficiency is below 60\%.

\begin{table}[htb!]
\renewcommand{\arraystretch}{1.3}
\begin{tabular}{|c|c|c|c|c|c|c|}
\hline \hline
\multirow{2}{*}{$p_T^W > $} & \multicolumn{2}{c|}{Filtering} & \multicolumn{2}{c|}{Pruning} & \multicolumn{2}{c|}{$J_{E_T}^{\rm II}$} \\ \cline{2-7}
                  & $R$(C/A)            & $\;R_{\rm filter}\;$          & $R$(anti-$k_t$)          & $\;\;\;z_\text{cut}\;\;\;$    &$\;\;\;\beta\;\;\;$ & $\;\;\;\gamma\;\;\;$      \\ \hline \hline
$\;200$~GeV\;               & 1.6          & \;0.35\;           & 1.0         & 0.075         &6 &5\\ \hline
$\;300$~GeV\;               & 1.4          & \;0.30\;           & 0.8         & 0.05         &8.5 &7\\ \hline
$\;400$~GeV\;              & 1.0          & \;0.25\;           & 0.8         & 0.05           &11 &9.5\\ \hline
$\;800$~GeV\;              & 0.6          & \;0.15\;           & 0.6         & 0.025          &12 & 7\\ \hline \hline
\end{tabular}
\caption{The optimized jet-finding parameters that provide the maximal signal efficiency for different $p_T$ cuts in the jet mass window of $(60, 100)$~GeV. For pruning, we start from an anti-$k_t$ fat jet and prune it using the Cambridge/Aachen (C/A)  algorithm. The additional parameter $D_\text{cut}$ is fixed to be $D_\text{cut} = m_J/p_{T_J}$. A smaller value of $D_\text{cut}$ provides marginally larger signal efficiencies, but with the signal peak shifted to a lower value.
}
\label{tab:params} 
\end{table}

To test our jet-finding algorithm for $W$'s with different $p_T$'s, we scan the jet parameters to obtain the maximal signal efficiency and show the values of the jet parameters in Table~\ref{tab:params}. For the fat jet plus pruning method, we choose to start with anti-$k_t$ fat jets, which give us slightly better results than starting from Cambridge/Aachen (C/A)~\cite{Dokshitzer:1997in,Wobisch:1998wt} fat jets, and then prune them using the C/A algorithm. For the fat jet plus filtering method, we use the C/A algorithm for both the fat jet and the filtering procedure. We have found that including only two subjets provides a larger signal efficiency in the jet mass window of $(60, 100)$~GeV, than three subjets. For our $J_{E_T}^{\rm II}$ algorithm, larger values of $\beta$ and $\beta-\gamma$ are preferred for $W$'s with a larger $p_T$. 

\begin{figure}[bht!]
    \centering
    \includegraphics[width=8cm]{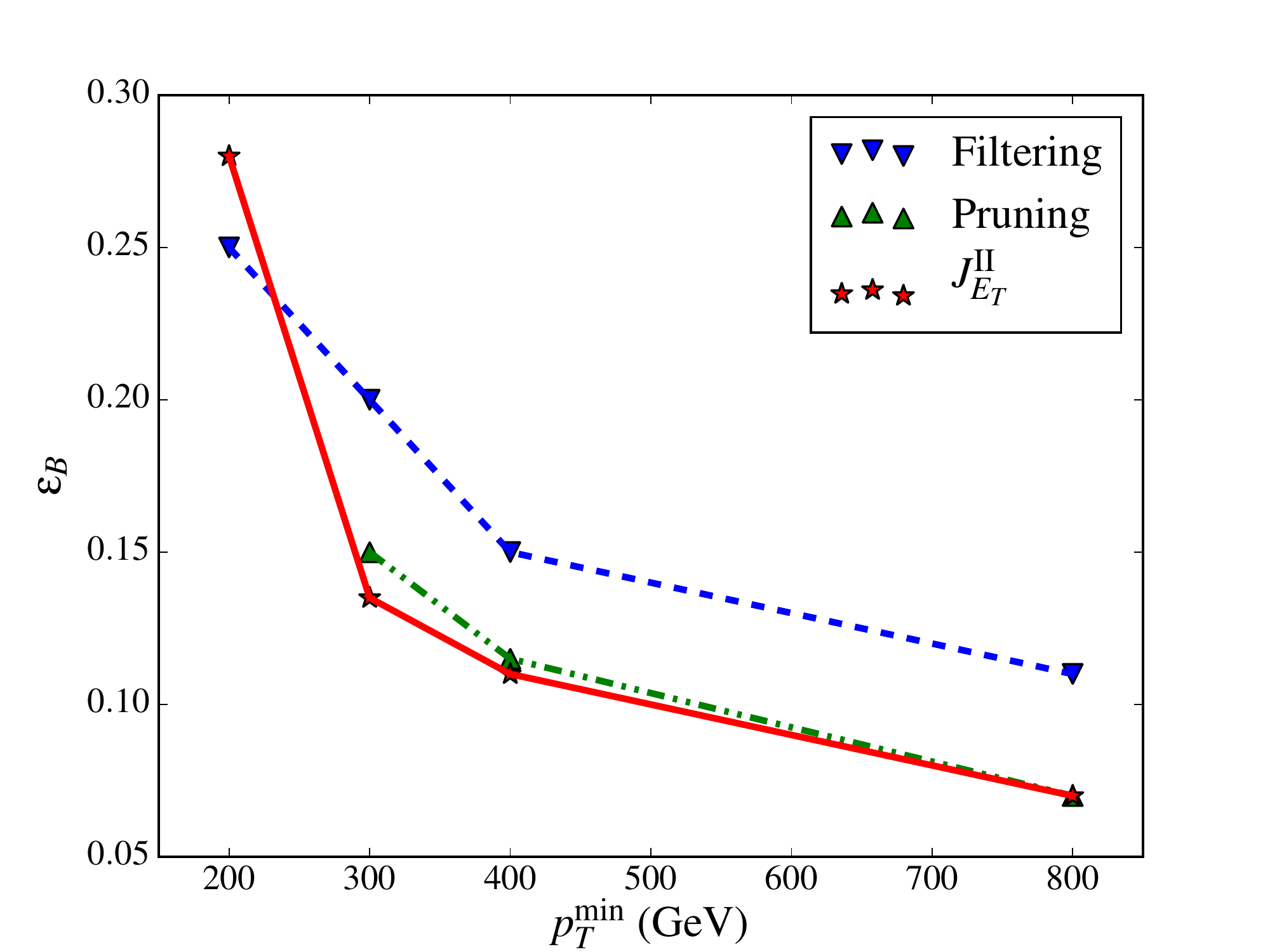}
    \caption{The QCD dijet background fake rate for different $p_T$ cuts with a fixed signal efficiency at 70\%. The jet mass window is chosen to be $(60, 100)$~GeV. The pruning case with $p_T > p_T^{\rm min} = 200$~GeV is not shown because the maximal signal efficiency that it can reach is only around 60\%.  }
    \label{fig:pt}
\end{figure}

After fixing the jet parameters for different $p_T$ cuts, we show the background fake rate in Fig.~\ref{fig:pt} with a fixed signal efficiency of 70\%. From Fig.~\ref{fig:pt}, one can see that for less boosted $W$'s with a small $p_T$ cut, the filtering method better identifies the two isolated jets and does a slightly better job in rejecting the background. For very boosted $W$'s, our $J_{E_T}^{\rm II}$ and the pruning methods provide comparable results, which are better than the filtering method because they do not force two separate prongs as the filtering does. For the intermediate $p_T$ cuts, our $J_{E_T}^{\rm II}$ algorithm has smaller background fake rates than the other two methods. Therefore, one can use our method to more efficiently tag hadronic $W$-jets for a wide range of $p_T$'s. 

\noindent
{\it{\textbf{Results for Heavy Diboson Resonance.}}}
To further illustrate the performance of our algorithm, we consider a beyond-the-Standard-Model-physics example, motivated by the recent ATLAS searches~\cite{Aad:2015owa} for diboson resonances using jet-substructure techniques. We consider a heavy $W^\prime$ gauge boson in the model of Ref.~\cite{Altarelli:1989ff}, produced at the 8 TeV LHC with 20 fb${}^{-1}$ integrated luminosity. We fix the $W^\prime$ mass to be 2 TeV and let it decay to $W+Z$, both of which decay hadronically.

\begin{figure}[thb!]
    \centering
    \includegraphics[width=0.235\textwidth, clip=true,viewport= 10 0 530 400]{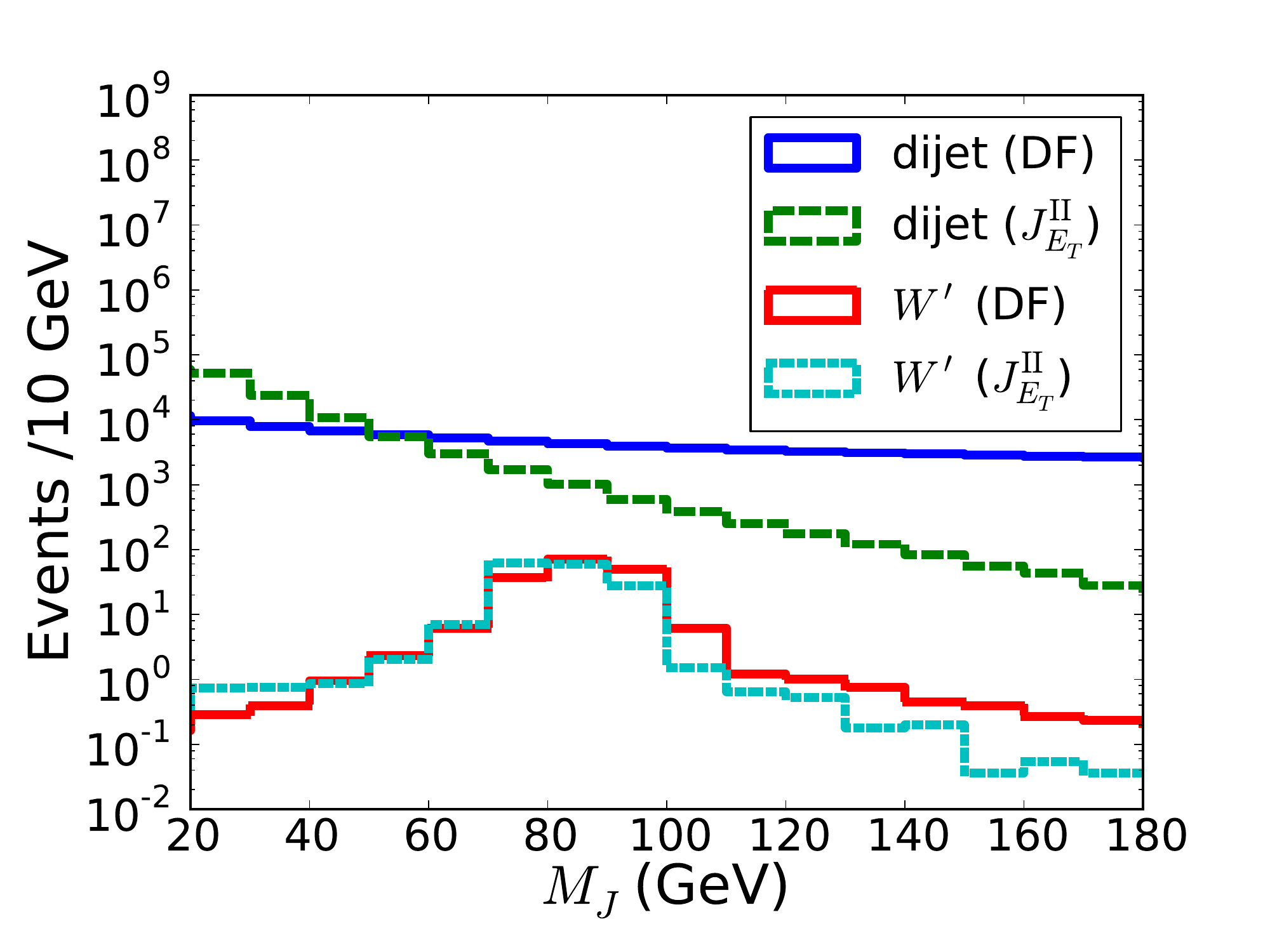}
    \includegraphics[width=0.235\textwidth, clip=true,viewport= 10 0 530 400]{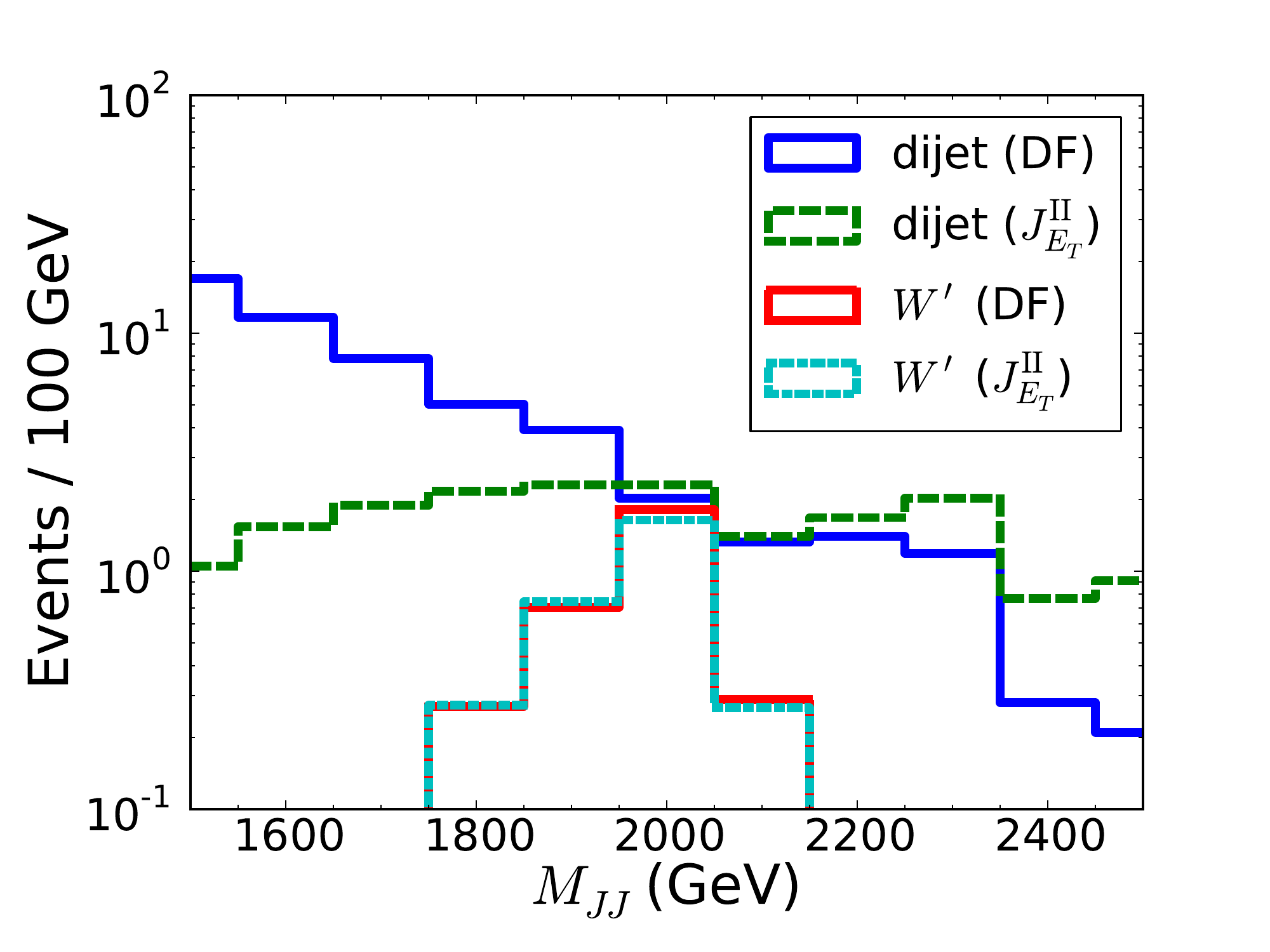}
    \caption{Left panel: the signal and background distributions for the leading two jet masses from our $J_{E_T}^{\rm II}$ and declustering-filtering (DF) procedures. Right panel: the distributions for the invariant masses of two leading jets with the lower(higher) jet mass within a 26 GeV mass window centered at the $W(Z)$ mass.}
    \label{fig:diboson}
\end{figure}

Following the ATLAS analysis in Ref.~\cite{Aad:2015owa}, we also use the declustering-filtering (DF) procedure to identify the $W$ and $Z$ bosons and to keep the QCD dijet mistag rate to be around 1\%. Starting from $R=1.2$ C/A fat jet with $p_T > 540$~GeV, we impose the same cuts on the mass-drop, subjet momentum balance and filtering parameters with $\mu_f =1$, $\sqrt{y_f}=0.2$ and $R_r =0.3$.~ \footnote{Our goal is not to exactly reproduce ATLAS results, which contain a detector simulation. For instance, we do not impose the number-of-track cut in our analysis, which could be more sensitive to detailed pileup information.} For our $J_{E_T}^{\rm II}$ jet-finding algorithm ($\beta=100$ and $\gamma=92$) and to further suppress the QCD background, we impose an additional cut on the momentum balance to utilize the fact that a QCD jet has more radiation around its axis than a $W/Z$ jet: in each fat jet, we construct the subleading $J_{E_T}^{\rm II}$ jet around the leading one and require its transverse momentum to be smaller than 0.015 times that of the leading one.~\footnote{One could also design a ``telescoping" jet cut~\cite{Chien:2013kca} by choosing a different set of $\beta$ and $\gamma$.}

After selecting the two leading jets with the lower(higher) jet mass within a 26 GeV mass window centered at the $W$($Z$) mass, we require them to have $|y_1-y_2|<1.2$ and $(p_{T_1}-p_{T_2})/(p_{T_1}+p_{T_2}) < 0.15$ like in the ATLAS analysis. We show the distributions of the individual jet mass in the left panel of Fig.~\ref{fig:diboson} and the dijet invariant mass distributions in the right panel. The left panel of Fig.~\ref{fig:diboson} clearly shows that our $J_{E_T}^{\rm II}$ algorithm is more efficient than the DF method to reduce QCD dijet events with a jet mass within the $W/Z$ boson mass window. From the right panel and imposing an invariant mass window cut of $(1.8, 2.2)$~TeV, our $J_{E_T}^{\rm II}$ method has improved $S/\sqrt{B}$ by 15\% compared to the DF method. We also note here that the QCD dijet invariant mass distributions from our method has a harder tail and a bump around 1.9 TeV. This can be understood from the basic properties of our jet function in Eq.~(\ref{eq:jet-hadron}), which prefers jets with a small jet mass but a large $p_T$. The bump location is generated from the basic $p_T$ cut on the leading two jets and the additional cut on the surrounding subleading jet $p_T$.  

\noindent
{\it{\textbf{Discussion and Summary.}}}
Our jet-finding algorithm has the advantage of identifying a smaller set of particles relevant for the boosted two-prong objects. On the other hand, it cannot capture other properties of boosted objects like the number of charged particles, the momentum balance of subjets or other color information of the jet~\cite{Krohn:2012fg,Cui:2010km}.  One should also apply the existing jet-substructure variables to our $J_{E_T}^{\rm II}$ jets to further enhance the new physics discovery potential. 

One could simply apply the $J_{E_T}^{\rm II}$ algorithm to identify the boosted Higgs boson with $h \rightarrow b\bar{b}$. For the single $b$-tagging case, one can first identify a two-prong jet and then check the additional $b$-tagging information for particles belong to this jet. For the double $b$-tagging case, $h\rightarrow b\bar{b}$ already guarantees the two-prong structure and our $J_{E_T}^{\rm II}$ may not add any additional discrimination power. It is interesting to see how to extend our $J_{E_T}^{\rm II}$ algorithm together with the $b$-tagging method. 

Beyond the two-prong jets, one could follow the general philosophy in this paper to identify a jet function favoring other decay topologies. There is not a single Fox-Wolfram moment that favors a three-prong jet like a hadronic top. On the other hand, it is still possible to define a jet function using a series of Fox-Wolfram moments, or other functions constructed to match the special event shape~\cite{Fox:1979id}. 

In conclusion, we have introduced a new jet-finding algorithm that favors two-prong jets. For the Standard Model $WW$ productions with two hadronic $W$'s, our $J_{E_T}^{\rm II}$ algorithm provides performance close to or better than the results from traditional jet grooming methods like filtering and pruning for a wide range of $p_{T}$'s. For a heavy di-boson resonance decaying to two hadronic weak gauge bosons, our $J_{E_T}^{\rm II}$ algorithm can further improve the new particle discovery significance.

\vspace{2mm}
\noindent
{\it{\textbf{Acknowledgments.}}}
We would like to thank Vernon Barger, Steve Ellis, Jesse Thaler and Lian-Tao Wang for useful discussion and comments. Y.~Bai and R.~Lu are supported by the U. S. Department of Energy under the contract DE-FG-02-95ER40896. Z.~Han is in part supported by U.~S.~Department of Energy under grant numbers DE-FG02-96ER40969 and DE-FG02-13ER41986.
\bibliography{w}

\begin{thebibliography}{30}%
\makeatletter
\providecommand \@ifxundefined [1]{%
 \@ifx{#1\undefined}
}%
\providecommand \@ifnum [1]{%
 \ifnum #1\expandafter \@firstoftwo
 \else \expandafter \@secondoftwo
 \fi
}%
\providecommand \@ifx [1]{%
 \ifx #1\expandafter \@firstoftwo
 \else \expandafter \@secondoftwo
 \fi
}%
\providecommand \natexlab [1]{#1}%
\providecommand \enquote  [1]{``#1''}%
\providecommand \bibnamefont  [1]{#1}%
\providecommand \bibfnamefont [1]{#1}%
\providecommand \citenamefont [1]{#1}%
\providecommand \href@noop [0]{\@secondoftwo}%
\providecommand \href [0]{\begingroup \@sanitize@url \@href}%
\providecommand \@href[1]{\@@startlink{#1}\@@href}%
\providecommand \@@href[1]{\endgroup#1\@@endlink}%
\providecommand \@sanitize@url [0]{\catcode `\\12\catcode `\$12\catcode
  `\&12\catcode `\#12\catcode `\^12\catcode `\_12\catcode `\%12\relax}%
\providecommand \@@startlink[1]{}%
\providecommand \@@endlink[0]{}%
\providecommand \url  [0]{\begingroup\@sanitize@url \@url }%
\providecommand \@url [1]{\endgroup\@href {#1}{\urlprefix }}%
\providecommand \urlprefix  [0]{URL }%
\providecommand \Eprint [0]{\href }%
\providecommand \doibase [0]{http://dx.doi.org/}%
\providecommand \selectlanguage [0]{\@gobble}%
\providecommand \bibinfo  [0]{\@secondoftwo}%
\providecommand \bibfield  [0]{\@secondoftwo}%
\providecommand \translation [1]{[#1]}%
\providecommand \BibitemOpen [0]{}%
\providecommand \bibitemStop [0]{}%
\providecommand \bibitemNoStop [0]{.\EOS\space}%
\providecommand \EOS [0]{\spacefactor3000\relax}%
\providecommand \BibitemShut  [1]{\csname bibitem#1\endcsname}%
\let\auto@bib@innerbib\@empty
\bibitem [{\citenamefont {Ellis}\ and\ \citenamefont
  {Soper}(1993)}]{Ellis:1993tq}%
  \BibitemOpen
  \bibfield  {author} {\bibinfo {author} {\bibfnamefont {S.~D.}\ \bibnamefont
  {Ellis}}\ and\ \bibinfo {author} {\bibfnamefont {D.~E.}\ \bibnamefont
  {Soper}},\ }\href {\doibase 10.1103/PhysRevD.48.3160} {\bibfield  {journal}
  {\bibinfo  {journal} {Phys. Rev.}\ }\textbf {\bibinfo {volume} {D48}},\
  \bibinfo {pages} {3160} (\bibinfo {year} {1993})},\ \Eprint
  {http://arxiv.org/abs/hep-ph/9305266} {arXiv:hep-ph/9305266 [hep-ph]}
  \BibitemShut {NoStop}%
\bibitem [{\citenamefont {Salam}(2010)}]{Salam:2009jx}%
  \BibitemOpen
  \bibfield  {author} {\bibinfo {author} {\bibfnamefont {G.~P.}\ \bibnamefont
  {Salam}},\ }\href {\doibase 10.1140/epjc/s10052-010-1314-6} {\bibfield
  {journal} {\bibinfo  {journal} {Eur. Phys. J.}\ }\textbf {\bibinfo {volume}
  {C67}},\ \bibinfo {pages} {637} (\bibinfo {year} {2010})},\ \Eprint
  {http://arxiv.org/abs/0906.1833} {arXiv:0906.1833 [hep-ph]} \BibitemShut
  {NoStop}%
\bibitem [{\citenamefont {Butterworth}\ \emph {et~al.}(2008)\citenamefont
  {Butterworth}, \citenamefont {Davison}, \citenamefont {Rubin},\ and\
  \citenamefont {Salam}}]{Butterworth:2008iy}%
  \BibitemOpen
  \bibfield  {author} {\bibinfo {author} {\bibfnamefont {J.~M.}\ \bibnamefont
  {Butterworth}}, \bibinfo {author} {\bibfnamefont {A.~R.}\ \bibnamefont
  {Davison}}, \bibinfo {author} {\bibfnamefont {M.}~\bibnamefont {Rubin}}, \
  and\ \bibinfo {author} {\bibfnamefont {G.~P.}\ \bibnamefont {Salam}},\ }\href
  {\doibase 10.1103/PhysRevLett.100.242001} {\bibfield  {journal} {\bibinfo
  {journal} {Phys.Rev.Lett.}\ }\textbf {\bibinfo {volume} {100}},\ \bibinfo
  {pages} {242001} (\bibinfo {year} {2008})},\ \Eprint
  {http://arxiv.org/abs/0802.2470} {arXiv:0802.2470 [hep-ph]} \BibitemShut
  {NoStop}%
\bibitem [{\citenamefont {Ellis}\ \emph {et~al.}(2009)\citenamefont {Ellis},
  \citenamefont {Vermilion},\ and\ \citenamefont {Walsh}}]{Ellis:2009su}%
  \BibitemOpen
  \bibfield  {author} {\bibinfo {author} {\bibfnamefont {S.~D.}\ \bibnamefont
  {Ellis}}, \bibinfo {author} {\bibfnamefont {C.~K.}\ \bibnamefont
  {Vermilion}}, \ and\ \bibinfo {author} {\bibfnamefont {J.~R.}\ \bibnamefont
  {Walsh}},\ }\href {\doibase 10.1103/PhysRevD.80.051501} {\bibfield  {journal}
  {\bibinfo  {journal} {Phys. Rev.}\ }\textbf {\bibinfo {volume} {D80}},\
  \bibinfo {pages} {051501} (\bibinfo {year} {2009})},\ \Eprint
  {http://arxiv.org/abs/0903.5081} {arXiv:0903.5081 [hep-ph]} \BibitemShut
  {NoStop}%
\bibitem [{\citenamefont {Ellis}\ \emph {et~al.}(2010)\citenamefont {Ellis},
  \citenamefont {Vermilion},\ and\ \citenamefont {Walsh}}]{Ellis:2009me}%
  \BibitemOpen
  \bibfield  {author} {\bibinfo {author} {\bibfnamefont {S.~D.}\ \bibnamefont
  {Ellis}}, \bibinfo {author} {\bibfnamefont {C.~K.}\ \bibnamefont
  {Vermilion}}, \ and\ \bibinfo {author} {\bibfnamefont {J.~R.}\ \bibnamefont
  {Walsh}},\ }\href {\doibase 10.1103/PhysRevD.81.094023} {\bibfield  {journal}
  {\bibinfo  {journal} {Phys. Rev.}\ }\textbf {\bibinfo {volume} {D81}},\
  \bibinfo {pages} {094023} (\bibinfo {year} {2010})},\ \Eprint
  {http://arxiv.org/abs/0912.0033} {arXiv:0912.0033 [hep-ph]} \BibitemShut
  {NoStop}%
\bibitem [{\citenamefont {Krohn}\ \emph {et~al.}(2010)\citenamefont {Krohn},
  \citenamefont {Thaler},\ and\ \citenamefont {Wang}}]{Krohn:2009th}%
  \BibitemOpen
  \bibfield  {author} {\bibinfo {author} {\bibfnamefont {D.}~\bibnamefont
  {Krohn}}, \bibinfo {author} {\bibfnamefont {J.}~\bibnamefont {Thaler}}, \
  and\ \bibinfo {author} {\bibfnamefont {L.-T.}\ \bibnamefont {Wang}},\ }\href
  {\doibase 10.1007/JHEP02(2010)084} {\bibfield  {journal} {\bibinfo  {journal}
  {JHEP}\ }\textbf {\bibinfo {volume} {02}},\ \bibinfo {pages} {084} (\bibinfo
  {year} {2010})},\ \Eprint {http://arxiv.org/abs/0912.1342} {arXiv:0912.1342
  [hep-ph]} \BibitemShut {NoStop}%
\bibitem [{\citenamefont {Larkoski}\ \emph {et~al.}(2014)\citenamefont
  {Larkoski}, \citenamefont {Marzani}, \citenamefont {Soyez},\ and\
  \citenamefont {Thaler}}]{Larkoski:2014wba}%
  \BibitemOpen
  \bibfield  {author} {\bibinfo {author} {\bibfnamefont {A.~J.}\ \bibnamefont
  {Larkoski}}, \bibinfo {author} {\bibfnamefont {S.}~\bibnamefont {Marzani}},
  \bibinfo {author} {\bibfnamefont {G.}~\bibnamefont {Soyez}}, \ and\ \bibinfo
  {author} {\bibfnamefont {J.}~\bibnamefont {Thaler}},\ }\href {\doibase
  10.1007/JHEP05(2014)146} {\bibfield  {journal} {\bibinfo  {journal} {JHEP}\
  }\textbf {\bibinfo {volume} {05}},\ \bibinfo {pages} {146} (\bibinfo {year}
  {2014})},\ \Eprint {http://arxiv.org/abs/1402.2657} {arXiv:1402.2657
  [hep-ph]} \BibitemShut {NoStop}%
\bibitem [{\citenamefont {Stewart}\ \emph {et~al.}(2010)\citenamefont
  {Stewart}, \citenamefont {Tackmann},\ and\ \citenamefont
  {Waalewijn}}]{Stewart:2010tn}%
  \BibitemOpen
  \bibfield  {author} {\bibinfo {author} {\bibfnamefont {I.~W.}\ \bibnamefont
  {Stewart}}, \bibinfo {author} {\bibfnamefont {F.~J.}\ \bibnamefont
  {Tackmann}}, \ and\ \bibinfo {author} {\bibfnamefont {W.~J.}\ \bibnamefont
  {Waalewijn}},\ }\href {\doibase 10.1103/PhysRevLett.105.092002} {\bibfield
  {journal} {\bibinfo  {journal} {Phys. Rev. Lett.}\ }\textbf {\bibinfo
  {volume} {105}},\ \bibinfo {pages} {092002} (\bibinfo {year} {2010})},\
  \Eprint {http://arxiv.org/abs/1004.2489} {arXiv:1004.2489 [hep-ph]}
  \BibitemShut {NoStop}%
\bibitem [{\citenamefont {Stewart}\ \emph {et~al.}(2015)\citenamefont
  {Stewart}, \citenamefont {Tackmann}, \citenamefont {Thaler}, \citenamefont
  {Vermilion},\ and\ \citenamefont {Wilkason}}]{Stewart:2015waa}%
  \BibitemOpen
  \bibfield  {author} {\bibinfo {author} {\bibfnamefont {I.~W.}\ \bibnamefont
  {Stewart}}, \bibinfo {author} {\bibfnamefont {F.~J.}\ \bibnamefont
  {Tackmann}}, \bibinfo {author} {\bibfnamefont {J.}~\bibnamefont {Thaler}},
  \bibinfo {author} {\bibfnamefont {C.~K.}\ \bibnamefont {Vermilion}}, \ and\
  \bibinfo {author} {\bibfnamefont {T.~F.}\ \bibnamefont {Wilkason}},\
  }\href@noop {} {\  (\bibinfo {year} {2015})},\ \Eprint
  {http://arxiv.org/abs/1508.01516} {arXiv:1508.01516 [hep-ph]} \BibitemShut
  {NoStop}%
\bibitem [{\citenamefont {Thaler}\ and\ \citenamefont
  {Wilkason}(2015)}]{Thaler:2015xaa}%
  \BibitemOpen
  \bibfield  {author} {\bibinfo {author} {\bibfnamefont {J.}~\bibnamefont
  {Thaler}}\ and\ \bibinfo {author} {\bibfnamefont {T.~F.}\ \bibnamefont
  {Wilkason}},\ }\href@noop {} {\  (\bibinfo {year} {2015})},\ \Eprint
  {http://arxiv.org/abs/1508.01518} {arXiv:1508.01518 [hep-ph]} \BibitemShut
  {NoStop}%
\bibitem [{\citenamefont {Georgi}(2014)}]{Georgi:2014zwa}%
  \BibitemOpen
  \bibfield  {author} {\bibinfo {author} {\bibfnamefont {H.}~\bibnamefont
  {Georgi}},\ }\href@noop {} {\  (\bibinfo {year} {2014})},\ \Eprint
  {http://arxiv.org/abs/1408.1161} {arXiv:1408.1161 [hep-ph]} \BibitemShut
  {NoStop}%
\bibitem [{\citenamefont {Ge}(2015)}]{Ge:2014ova}%
  \BibitemOpen
  \bibfield  {author} {\bibinfo {author} {\bibfnamefont {S.-F.}\ \bibnamefont
  {Ge}},\ }\href {\doibase 10.1007/JHEP05(2015)066} {\bibfield  {journal}
  {\bibinfo  {journal} {JHEP}\ }\textbf {\bibinfo {volume} {1505}},\ \bibinfo
  {pages} {066} (\bibinfo {year} {2015})},\ \Eprint
  {http://arxiv.org/abs/1408.3823} {arXiv:1408.3823 [hep-ph]} \BibitemShut
  {NoStop}%
\bibitem [{\citenamefont {Bai}\ \emph {et~al.}(2015)\citenamefont {Bai},
  \citenamefont {Han},\ and\ \citenamefont {Lu}}]{Bai:2014qca}%
  \BibitemOpen
  \bibfield  {author} {\bibinfo {author} {\bibfnamefont {Y.}~\bibnamefont
  {Bai}}, \bibinfo {author} {\bibfnamefont {Z.}~\bibnamefont {Han}}, \ and\
  \bibinfo {author} {\bibfnamefont {R.}~\bibnamefont {Lu}},\ }\href {\doibase
  10.1007/JHEP03(2015)102} {\bibfield  {journal} {\bibinfo  {journal} {JHEP}\
  }\textbf {\bibinfo {volume} {1503}},\ \bibinfo {pages} {102} (\bibinfo {year}
  {2015})},\ \Eprint {http://arxiv.org/abs/1411.3705} {arXiv:1411.3705
  [hep-ph]} \BibitemShut {NoStop}%
\bibitem [{\citenamefont {Kaufmann}\ \emph {et~al.}(2015)\citenamefont
  {Kaufmann}, \citenamefont {Mukherjee},\ and\ \citenamefont
  {Vogelsang}}]{Kaufmann:2014nda}%
  \BibitemOpen
  \bibfield  {author} {\bibinfo {author} {\bibfnamefont {T.}~\bibnamefont
  {Kaufmann}}, \bibinfo {author} {\bibfnamefont {A.}~\bibnamefont {Mukherjee}},
  \ and\ \bibinfo {author} {\bibfnamefont {W.}~\bibnamefont {Vogelsang}},\
  }\href {\doibase 10.1103/PhysRevD.91.034001} {\bibfield  {journal} {\bibinfo
  {journal} {Phys.Rev.}\ }\textbf {\bibinfo {volume} {D91}},\ \bibinfo {pages}
  {034001} (\bibinfo {year} {2015})},\ \Eprint {http://arxiv.org/abs/1412.0298}
  {arXiv:1412.0298 [hep-ph]} \BibitemShut {NoStop}%
\bibitem [{\citenamefont {Thaler}(2015)}]{Thaler:2015uja}%
  \BibitemOpen
  \bibfield  {author} {\bibinfo {author} {\bibfnamefont {J.}~\bibnamefont
  {Thaler}},\ }\href@noop {} {\  (\bibinfo {year} {2015})},\ \Eprint
  {http://arxiv.org/abs/1506.07876} {arXiv:1506.07876 [hep-ph]} \BibitemShut
  {NoStop}%
\bibitem [{\citenamefont {Fox}\ and\ \citenamefont
  {Wolfram}(1978)}]{Fox:1978vu}%
  \BibitemOpen
  \bibfield  {author} {\bibinfo {author} {\bibfnamefont {G.~C.}\ \bibnamefont
  {Fox}}\ and\ \bibinfo {author} {\bibfnamefont {S.}~\bibnamefont {Wolfram}},\
  }\href {\doibase 10.1103/PhysRevLett.41.1581} {\bibfield  {journal} {\bibinfo
   {journal} {Phys.Rev.Lett.}\ }\textbf {\bibinfo {volume} {41}},\ \bibinfo
  {pages} {1581} (\bibinfo {year} {1978})}\BibitemShut {NoStop}%
\bibitem [{\citenamefont {Fox}\ and\ \citenamefont
  {Wolfram}(1979{\natexlab{a}})}]{Fox:1978vw}%
  \BibitemOpen
  \bibfield  {author} {\bibinfo {author} {\bibfnamefont {G.~C.}\ \bibnamefont
  {Fox}}\ and\ \bibinfo {author} {\bibfnamefont {S.}~\bibnamefont {Wolfram}},\
  }\href {\doibase 10.1016/0550-3213(79)90003-8} {\bibfield  {journal}
  {\bibinfo  {journal} {Nucl. Phys.}\ }\textbf {\bibinfo {volume} {B149}},\
  \bibinfo {pages} {413} (\bibinfo {year} {1979}{\natexlab{a}})},\ \bibinfo
  {note} {[Erratum: Nucl. Phys.B157,543(1979)]}\BibitemShut {NoStop}%
\bibitem [{\citenamefont {Cacciari}\ \emph
  {et~al.}(2008{\natexlab{a}})\citenamefont {Cacciari}, \citenamefont {Salam},\
  and\ \citenamefont {Soyez}}]{Cacciari:2008gn}%
  \BibitemOpen
  \bibfield  {author} {\bibinfo {author} {\bibfnamefont {M.}~\bibnamefont
  {Cacciari}}, \bibinfo {author} {\bibfnamefont {G.~P.}\ \bibnamefont {Salam}},
  \ and\ \bibinfo {author} {\bibfnamefont {G.}~\bibnamefont {Soyez}},\ }\href
  {\doibase 10.1088/1126-6708/2008/04/005} {\bibfield  {journal} {\bibinfo
  {journal} {JHEP}\ }\textbf {\bibinfo {volume} {0804}},\ \bibinfo {pages}
  {005} (\bibinfo {year} {2008}{\natexlab{a}})},\ \Eprint
  {http://arxiv.org/abs/0802.1188} {arXiv:0802.1188 [hep-ph]} \BibitemShut
  {NoStop}%
\bibitem [{\citenamefont {Cacciari}\ \emph
  {et~al.}(2008{\natexlab{b}})\citenamefont {Cacciari}, \citenamefont {Salam},\
  and\ \citenamefont {Soyez}}]{Cacciari:2008gp}%
  \BibitemOpen
  \bibfield  {author} {\bibinfo {author} {\bibfnamefont {M.}~\bibnamefont
  {Cacciari}}, \bibinfo {author} {\bibfnamefont {G.~P.}\ \bibnamefont {Salam}},
  \ and\ \bibinfo {author} {\bibfnamefont {G.}~\bibnamefont {Soyez}},\ }\href
  {\doibase 10.1088/1126-6708/2008/04/063} {\bibfield  {journal} {\bibinfo
  {journal} {JHEP}\ }\textbf {\bibinfo {volume} {04}},\ \bibinfo {pages} {063}
  (\bibinfo {year} {2008}{\natexlab{b}})},\ \Eprint
  {http://arxiv.org/abs/0802.1189} {arXiv:0802.1189 [hep-ph]} \BibitemShut
  {NoStop}%
\bibitem [{\citenamefont {Cacciari}\ \emph {et~al.}(2012)\citenamefont
  {Cacciari}, \citenamefont {Salam},\ and\ \citenamefont
  {Soyez}}]{Cacciari:2011ma}%
  \BibitemOpen
  \bibfield  {author} {\bibinfo {author} {\bibfnamefont {M.}~\bibnamefont
  {Cacciari}}, \bibinfo {author} {\bibfnamefont {G.~P.}\ \bibnamefont {Salam}},
  \ and\ \bibinfo {author} {\bibfnamefont {G.}~\bibnamefont {Soyez}},\ }\href
  {\doibase 10.1140/epjc/s10052-012-1896-2} {\bibfield  {journal} {\bibinfo
  {journal} {Eur.Phys.J.}\ }\textbf {\bibinfo {volume} {C72}},\ \bibinfo
  {pages} {1896} (\bibinfo {year} {2012})},\ \Eprint
  {http://arxiv.org/abs/1111.6097} {arXiv:1111.6097 [hep-ph]} \BibitemShut
  {NoStop}%
\bibitem [{\citenamefont {Sjöstrand}\ \emph {et~al.}(2015)\citenamefont
  {Sjöstrand}, \citenamefont {Ask}, \citenamefont {Christiansen},
  \citenamefont {Corke}, \citenamefont {Desai} \emph
  {et~al.}}]{Sjostrand:2014zea}%
  \BibitemOpen
  \bibfield  {author} {\bibinfo {author} {\bibfnamefont {T.}~\bibnamefont
  {Sjöstrand}}, \bibinfo {author} {\bibfnamefont {S.}~\bibnamefont {Ask}},
  \bibinfo {author} {\bibfnamefont {J.~R.}\ \bibnamefont {Christiansen}},
  \bibinfo {author} {\bibfnamefont {R.}~\bibnamefont {Corke}}, \bibinfo
  {author} {\bibfnamefont {N.}~\bibnamefont {Desai}},  \emph {et~al.},\ }\href
  {\doibase 10.1016/j.cpc.2015.01.024} {\bibfield  {journal} {\bibinfo
  {journal} {Comput.Phys.Commun.}\ }\textbf {\bibinfo {volume} {191}},\
  \bibinfo {pages} {159} (\bibinfo {year} {2015})},\ \Eprint
  {http://arxiv.org/abs/1410.3012} {arXiv:1410.3012 [hep-ph]} \BibitemShut
  {NoStop}%
\bibitem [{\citenamefont {Thaler}\ and\ \citenamefont
  {Van~Tilburg}(2011)}]{Thaler:2010tr}%
  \BibitemOpen
  \bibfield  {author} {\bibinfo {author} {\bibfnamefont {J.}~\bibnamefont
  {Thaler}}\ and\ \bibinfo {author} {\bibfnamefont {K.}~\bibnamefont
  {Van~Tilburg}},\ }\href {\doibase 10.1007/JHEP03(2011)015} {\bibfield
  {journal} {\bibinfo  {journal} {JHEP}\ }\textbf {\bibinfo {volume} {03}},\
  \bibinfo {pages} {015} (\bibinfo {year} {2011})},\ \Eprint
  {http://arxiv.org/abs/1011.2268} {arXiv:1011.2268 [hep-ph]} \BibitemShut
  {NoStop}%
\bibitem [{\citenamefont {Dokshitzer}\ \emph {et~al.}(1997)\citenamefont
  {Dokshitzer}, \citenamefont {Leder}, \citenamefont {Moretti},\ and\
  \citenamefont {Webber}}]{Dokshitzer:1997in}%
  \BibitemOpen
  \bibfield  {author} {\bibinfo {author} {\bibfnamefont {Y.~L.}\ \bibnamefont
  {Dokshitzer}}, \bibinfo {author} {\bibfnamefont {G.~D.}\ \bibnamefont
  {Leder}}, \bibinfo {author} {\bibfnamefont {S.}~\bibnamefont {Moretti}}, \
  and\ \bibinfo {author} {\bibfnamefont {B.~R.}\ \bibnamefont {Webber}},\
  }\href {\doibase 10.1088/1126-6708/1997/08/001} {\bibfield  {journal}
  {\bibinfo  {journal} {JHEP}\ }\textbf {\bibinfo {volume} {08}},\ \bibinfo
  {pages} {001} (\bibinfo {year} {1997})},\ \Eprint
  {http://arxiv.org/abs/hep-ph/9707323} {arXiv:hep-ph/9707323 [hep-ph]}
  \BibitemShut {NoStop}%
\bibitem [{\citenamefont {Wobisch}\ and\ \citenamefont
  {Wengler}(1998)}]{Wobisch:1998wt}%
  \BibitemOpen
  \bibfield  {author} {\bibinfo {author} {\bibfnamefont {M.}~\bibnamefont
  {Wobisch}}\ and\ \bibinfo {author} {\bibfnamefont {T.}~\bibnamefont
  {Wengler}}\ }(\bibinfo {year} {1998})\ \Eprint
  {http://arxiv.org/abs/hep-ph/9907280} {arXiv:hep-ph/9907280 [hep-ph]}
  \BibitemShut {NoStop}%
\bibitem [{\citenamefont {Aad}\ \emph {et~al.}(2015)\citenamefont {Aad} \emph
  {et~al.}}]{Aad:2015owa}%
  \BibitemOpen
  \bibfield  {author} {\bibinfo {author} {\bibfnamefont {G.}~\bibnamefont
  {Aad}} \emph {et~al.} (\bibinfo {collaboration} {ATLAS}),\ }\href@noop {} {\
  (\bibinfo {year} {2015})},\ \Eprint {http://arxiv.org/abs/1506.00962}
  {arXiv:1506.00962 [hep-ex]} \BibitemShut {NoStop}%
\bibitem [{\citenamefont {Altarelli}\ \emph {et~al.}(1989)\citenamefont
  {Altarelli}, \citenamefont {Mele},\ and\ \citenamefont
  {Ruiz-Altaba}}]{Altarelli:1989ff}%
  \BibitemOpen
  \bibfield  {author} {\bibinfo {author} {\bibfnamefont {G.}~\bibnamefont
  {Altarelli}}, \bibinfo {author} {\bibfnamefont {B.}~\bibnamefont {Mele}}, \
  and\ \bibinfo {author} {\bibfnamefont {M.}~\bibnamefont {Ruiz-Altaba}},\
  }\href {\doibase 10.1007/BF01552335, 10.1007/BF01556677} {\bibfield
  {journal} {\bibinfo  {journal} {Z. Phys.}\ }\textbf {\bibinfo {volume}
  {C45}},\ \bibinfo {pages} {109} (\bibinfo {year} {1989})},\ \bibinfo {note}
  {[Erratum: Z. Phys.C47,676(1990)]}\BibitemShut {NoStop}%
\bibitem [{\citenamefont {Chien}(2014)}]{Chien:2013kca}%
  \BibitemOpen
  \bibfield  {author} {\bibinfo {author} {\bibfnamefont {Y.-T.}\ \bibnamefont
  {Chien}},\ }\href {\doibase 10.1103/PhysRevD.90.054008} {\bibfield  {journal}
  {\bibinfo  {journal} {Phys. Rev.}\ }\textbf {\bibinfo {volume} {D90}},\
  \bibinfo {pages} {054008} (\bibinfo {year} {2014})},\ \Eprint
  {http://arxiv.org/abs/1304.5240} {arXiv:1304.5240 [hep-ph]} \BibitemShut
  {NoStop}%
\bibitem [{\citenamefont {Krohn}\ \emph {et~al.}(2013)\citenamefont {Krohn},
  \citenamefont {Schwartz}, \citenamefont {Lin},\ and\ \citenamefont
  {Waalewijn}}]{Krohn:2012fg}%
  \BibitemOpen
  \bibfield  {author} {\bibinfo {author} {\bibfnamefont {D.}~\bibnamefont
  {Krohn}}, \bibinfo {author} {\bibfnamefont {M.~D.}\ \bibnamefont {Schwartz}},
  \bibinfo {author} {\bibfnamefont {T.}~\bibnamefont {Lin}}, \ and\ \bibinfo
  {author} {\bibfnamefont {W.~J.}\ \bibnamefont {Waalewijn}},\ }\href {\doibase
  10.1103/PhysRevLett.110.212001} {\bibfield  {journal} {\bibinfo  {journal}
  {Phys. Rev. Lett.}\ }\textbf {\bibinfo {volume} {110}},\ \bibinfo {pages}
  {212001} (\bibinfo {year} {2013})},\ \Eprint {http://arxiv.org/abs/1209.2421}
  {arXiv:1209.2421 [hep-ph]} \BibitemShut {NoStop}%
\bibitem [{\citenamefont {Cui}\ \emph {et~al.}(2011)\citenamefont {Cui},
  \citenamefont {Han},\ and\ \citenamefont {Schwartz}}]{Cui:2010km}%
  \BibitemOpen
  \bibfield  {author} {\bibinfo {author} {\bibfnamefont {Y.}~\bibnamefont
  {Cui}}, \bibinfo {author} {\bibfnamefont {Z.}~\bibnamefont {Han}}, \ and\
  \bibinfo {author} {\bibfnamefont {M.~D.}\ \bibnamefont {Schwartz}},\ }\href
  {\doibase 10.1103/PhysRevD.83.074023} {\bibfield  {journal} {\bibinfo
  {journal} {Phys.Rev.}\ }\textbf {\bibinfo {volume} {D83}},\ \bibinfo {pages}
  {074023} (\bibinfo {year} {2011})},\ \Eprint {http://arxiv.org/abs/1012.2077}
  {arXiv:1012.2077 [hep-ph]} \BibitemShut {NoStop}%
\bibitem [{\citenamefont {Fox}\ and\ \citenamefont
  {Wolfram}(1979{\natexlab{b}})}]{Fox:1979id}%
  \BibitemOpen
  \bibfield  {author} {\bibinfo {author} {\bibfnamefont {G.~C.}\ \bibnamefont
  {Fox}}\ and\ \bibinfo {author} {\bibfnamefont {S.}~\bibnamefont {Wolfram}},\
  }\href {\doibase 10.1016/0370-2693(79)90444-1} {\bibfield  {journal}
  {\bibinfo  {journal} {Phys. Lett.}\ }\textbf {\bibinfo {volume} {B82}},\
  \bibinfo {pages} {134} (\bibinfo {year} {1979}{\natexlab{b}})}\BibitemShut
  {NoStop}%
\end{thebibliography}%
\bibliographystyle{apsrev4-1}
\end{document}